\documentstyle[twoside,fleqn,espcrc2,epsfig]{article}

\def \be {\begin{equation}}
\def \ee {\end{equation}}                                         
\def \ba {\begin{eqnarray}}
\def \ea {\end{eqnarray}}

\newcommand{\AmS}{{\protect\the\textfont2
  A\kern-.1667em\lower.5ex\hbox{M}\kern-.125emS}}

\title{On the SU(2)-Higgs Phase Transition}

\author{Isabel Campos\address{Departamento de F\'{\i}sica Te\'orica,  
        Universidad de Zaragoza,\\ Plaza San Francisco s/n, 
	50009 Zaragoza, Spain}%
        \thanks{Supported by a Fellowship from the Ministerio de Educaci\'on 
		y Cultura Espa\~nol.}}
       
\begin{document}

\begin{abstract}
We investigate the properties of the Confinement-Higgs phase transition in the 
SU(2)-Higgs model. 
The system is shown to exhibit a transient behavior up to {\it L}=24 
along which, the order of the phase transition cannot be discerned.
In order to get a global vision on the problem,
we have introduced a second (next-to-nearest neighbors) gauge-Higgs
coupling ($\kappa_2$). On this extended parameter
space we find a line of phase transitions becoming
increasely weak as the standard case is approached ($\kappa_2 = 0$).
From the global behavior on this parameter space we conclude
that the transition is also first order in the standard case. 
\end{abstract}

\maketitle

\section{INTRODUCTION}

The Higgs sector of the SM can be approximated by the SU(2)$\otimes$U(1)
Higgs model. If as a first approximation one  
neglects the U(1) gauge coupling
($g_{\mathrm{U}(1)} \approx \sin \theta_{W} g_{\mathrm{SU}(2)}$) 
one has to deal with the SU(2)-Higgs model:
\ba
S_{\lambda} & = & \displaystyle \beta \sum_{p} [1 - 
\frac{1}{2} {\mathrm{Tr}}U_p] +  
\sum_{x} \Phi^{\dagger}(x) \Phi(x)   \nonumber \\
& & \displaystyle  +
\lambda \sum_{x} [\Phi^{\dagger}(x) \Phi(x) - 1]^2    \nonumber \\
& & - \frac{1}{2} \kappa_1 \sum_{x, \mu} {\mathrm{Tr}}\Phi^{\dagger}(x) U_{\mu}(x) \Phi(x + \mu) 
\ea
This model was extensively studied in the eighties \cite{LRV,MONT}.
A phase transition (PT)
line separates a region where the scalar particles are confined
from another region where the spectrum consists of the W's bosons and the
Higgs particle. The PT ends at some finite point of the parameter
space, being both regions analytically connected.
Concerning the order of the PT line, the model exhibits a second
order PT with mean field exponents in the limit $\beta \rightarrow \infty$
,pure $\lambda \Phi^4$ theory. 
At the end-point the transition seems to be second order with mean
field exponents, \cite{ROB}, however, this issue has still to be clarified.
In the scaling region ($\beta \approx 2.5$) and for
small and intermediate values of $\lambda$ the PT is distinctly first
order, getting increasely weak as $\lambda$ is increased. In particular,
in the limit $\lambda = \infty$, which is equivalent to work with
fixed modulus of the Higgs field, the PT is commonly believed to 
be first order though extremely weak. 
However the numerical simulations up to date are not conclusive.

\section{THE MODEL}

To improve on the statistics we have first studied the model
with action $S_{\infty}$ up to {\it L} = 24, being the results compatible,
both, with a very weak first order PT and with a continuous one.
In order to get a more conclusive answer without going to prohibitive
large lattice sizes we have added a second positive gauge-Higgs coupling
$\kappa_2$ connecting next-to-nearest neighbors on the lattice.
The new action reads:
\ba
S & = & S_{\infty}
- \frac{1}{4} \kappa_2 \sum_{x, \mu<\nu}
{\rm {Tr}} \Phi^{\dagger}(x) [U_{\mu}(x) U_{\nu}(x+\mu)   \nonumber \\
& & + U_{\nu}(x) U_{\mu}(x+\nu)] \Phi(x + \mu + \nu)     
\label{ACTION}
\ea

\begin{table*}[hbt]
\setlength{\tabcolsep}{1.5pc}
\newlength{\digitwidth} \settowidth{\digitwidth}{\rm 0}
\catcode`?=\active \def?{\kern\digitwidth}
\caption{Latent heat and change in the action. The values at $\kappa_1$ = 0.3
are only upper bounds.}
\label{LAT}
\begin{tabular*}{\textwidth}{@{}l@{\extracolsep{\fill}}rrrr}
\hline
                 & \multicolumn{1}{r}{$\Delta E_1$} 
                 & \multicolumn{1}{r}{$\Delta E_2$} 
                 & \multicolumn{1}{r}{$\Delta S$}   \\
\hline
$\kappa_1$ = 0     & -  & $0.0366(8)$ & $0.0134(12)$ \\
$\kappa_1$ = 0.02  & $0.0162(6)$ & $0.0347(5)$ & $0.0137(13)$  \\
$\kappa_1$ = 0.1       & $0.0162(7)$ & $0.0345(9)$ & $0.0094(10)$  \\
$\kappa_1$ = 0.2        & $0.0179(7)$ & $0.0201(8)$ & $0.0078(12)$          \\
$\kappa_1$ = 0.3   & $\approx 0.006$ & $\approx 0.012$ & $\approx 0.0026$  \\

\hline

\end{tabular*}
\end{table*}

Heuristically speaking, 
the effect of the new coupling is to reinforce the PT,
but it should not change its order since the symmetry properties
remain unchanged. This coupling will also be used
as a parameter to monitorize the weakening mechanism of the PT. 

\section{RESULTS}

\begin{figure}[h]
\vglue 3mm
\epsfig{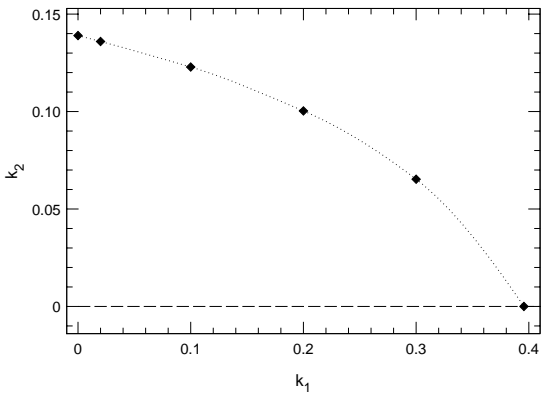}
\vglue -1cm
\caption{Phase Diagram obtained from the MC simulation with {\it L} = 12.}
\label{PhD}
\end{figure}

On the parameter space ($\beta,\kappa_1,\kappa_2$)
we consider the plane $\beta=2.3$. 
On this plane there is a PT line $\kappa_2^c(\kappa_1)$
along which, we shall study the existence of latent heat and the behavior
of the specific heat with the volume.
We worked at fixed values $\kappa_1$ = 0, 0.02, 0.1, 0.2, 0.3 and
$\kappa_2 = 0$ (standard case). To locate the critical line we sought
for the maximum of the specific heat ($C_v^{max}(L)$). We have used the
Spectral Density Method \cite{SDM} to locate the couplings where to measure.
The phase diagram we found with this
criterium is plotted in Fig. \ref{PhD} for {\it L} = 12.
 
\subsection{Latent Heat}

\begin{figure}[h]
\vglue 4mm
\epsfig{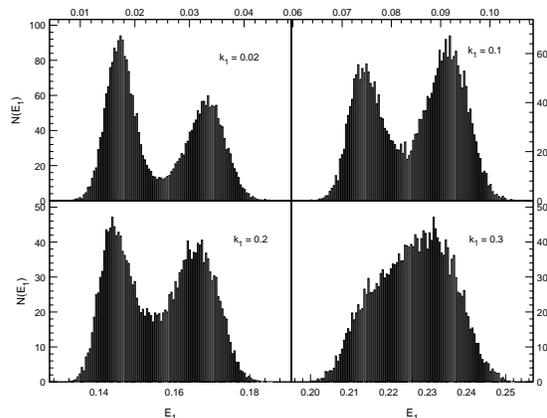}
\vglue -1cm
\caption{E$_1$ distributions for several values of $\kappa_1$
in {\it L} = 12.}
\label{HISTO}
\end{figure}

The PT line $\kappa_2^c(\kappa_1)$ turns out to be first order.
In Fig. \ref{HISTO} the energy distribution for {\it L} = 12
is plotted for $\kappa_1$ = 0.02, 0.1, 0.2 and 0.3. We observe
that the latent heat $\Delta E(L)$ is no longer measurable at
$\kappa_1$ = 0.3, and one has to go to
{\it L} = 20 lattice to see the first evidences about the existence
of latent heat at this value of the couplings. 
This evidences the weakening of the PT as the limit $\kappa_2$ = 0 
is approached. Weak first order PT appear
often in literature \cite{POTS} and they can be defined like first order PT
with small discontinuities. They are characterized by
a transient behavior during which the lattice size {\it L} is much smaller
than the correlation length at critical point, $\xi_c$. Along this regime
the PT transition behaves like a second order one since $\xi_c$ is
effectively infinite. Once the asymptotic regime is reached, 
{\it L} $\geq \xi_c$,
the first order character is evidenced appearing the typical two-peak
structure in the energy distribution.
In Table \ref{LAT} we report the results obtained for $\Delta E(\infty)$
by fitting the values of $\Delta E(L)$ to 1/L$^4$.

\begin{figure}[h]
\vglue 3mm
\epsfig{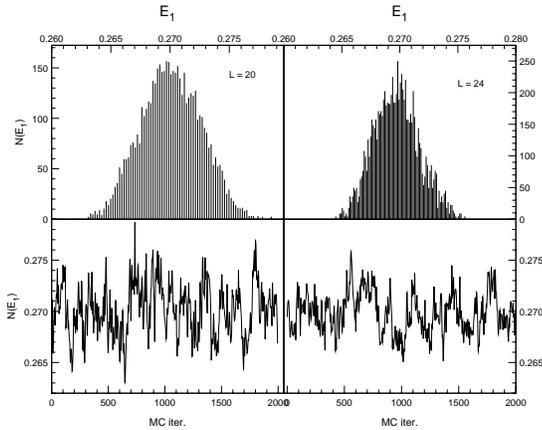}
\vglue -1cm
\caption{E$_1$ distributions and MC evolution at $\kappa_2$ = 0 in {\it L} = 20
and {\it L} = 24}
\label{K20}
\end{figure}

At $\kappa_2$ = 0 (standard case) we have no distinct observations
of two-peak energy distributions up to {\it L} = 24 (Fig. \ref{K20}).

\begin{figure}[h]
\vglue 3mm
\epsfig{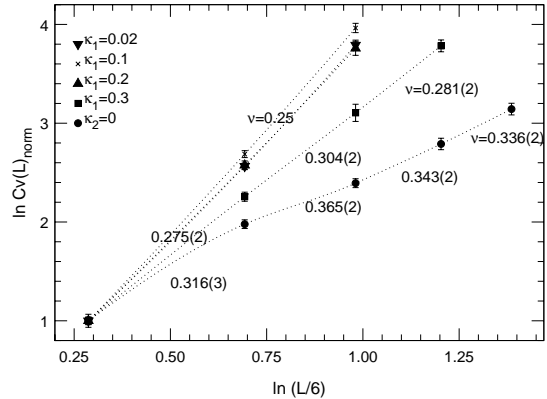}
\vglue -1cm
\caption{C$_v^{\it {max}}$ for the various $\kappa_1$ values and $\kappa_2=0$.
We have normalized the values with respect to 
C$_v^{\it {max}}$(8)/C$_v^{\it {max}}$(6).}
\label{CV}
\end{figure}

\subsection{Specific heat}

We now address the question of the behavior of the specific heat, 
$C_v({\it L})$ along the transient region of a weak first order
phase transition. We use, as a technical tool, a $pseudo$ $\nu$ exponent
in order to discern whether or not the asymptotic regime has been
reached since we expect to measure $\nu$ compatible with 0.25 (i.e. 1/d)
in the asymptotic region.
We have measured the evolution of $C_v({\it L})^{\it {max}}$ 
with the volume along the PT line. This maximum is expected to
scale with a $pseudo$ $\alpha/\nu$ index, and we obtain $\nu$
from the relation $\alpha = 2 - \nu d$.

In Fig. \ref{CV} we plot C$_v^{\it {max}}({\it L})$ relative to 
C$_v^{\it {max}}(6)$ as a function of the lattice size. 
The slope of the segment joining
the values of C$_v^{\it {max}}(\it L)$ in consecutive
lattices gives the $pseudo$ $\alpha/\nu$ exponent.
The values for $\nu$ are compatible with 0.25 at $\kappa_1$ = 0.02, 
0.1 and 0.2 for all the volumes we compare.  However the transition
at $\kappa_1$ = 0.3 seems to be much weak. We do not
have evidences of asymptoticity in C$_v^{\it {max}}(\it L)$ till L=20, as could
be expected from the energy distributions.
At $\kappa_2$ = 0, C$_v^{\it {max}}(\it L)$ behaves qualitatively in the
same way.

\section{CONCLUSIONS}

We have studied an extended parameter space in order to get
insight on the nature of the SU(2)-Higgs PT with $\lambda = \infty$
at T=0. We have found a line of first order phase transitions
getting increasely weak as the standard case ($\kappa_2$ = 0) 
is approached.
The results in this limit are compatible with a continuous PT
too. However, from the behavior in the global parameter space
of the specific heat we conclude that we are in the
transient region of a first order phase transition up to 
{\it L} = 24. 

I thank L.A. Fern\'andez and A. Taranc\'on for comments
and advice.

\end{document}